%% file: main.tex
\documentclass[10pt,twocolumn,letterpaper]{article}
\usepackage{iccv}             
\input{preamble}

\usepackage{graphicx}
\usepackage{amsmath}
\usepackage{amssymb}
\usepackage{booktabs}
\usepackage{multirow}
\usepackage{calc}
\usepackage{mathtools}
\definecolor{iccvblue}{rgb}{0.21,0.49,0.74}
\usepackage{hyperref}
\hypersetup{pagebackref,breaklinks,colorlinks,allcolors=iccvblue}
\usepackage[capitalize]{cleveref}
\makeatletter
\AtBeginDocument
 {
   \def\ltx@label#1{\cref@label{#1}}%add braces
   \def\label@in@display@noarg#1{\cref@old@label@in@display{#1}}%remove braces
\def\label@in@mmeasure@noarg#1{%
    \begingroup%
      \measuring@false%
      \cref@old@label@in@display{#1}%remove braces for multline, see https://tex.stackexchange.com/q/737204/2388
    \endgroup}%  
 } %
\makeatother
\usepackage[nolist]{acronym}
\usepackage[numbers]{natbib}
\usepackage{collect}
\usepackage{flushend}
\usepackage{adjustbox}
\usepackage{bm}
\usepackage{algorithm}
\usepackage{algpseudocode}
\usepackage{siunitx}
\usepackage{xspace}
\usepackage{collcell}
\usepackage{makecell}
\usepackage{comment}

\input{tr-commands}
\begin{document}
\begin{acronym}
\acro{AD}{automatic differentiation}
\acro{BRDF}{bi-directional reflectance distribution function}
\acro{CDF}{cumulative distribution function}
\acro{CNN}{convolutional neural network}
\acro{CG}{conjugate gradient}
\acro{FD}{finite differences}
\acro{GD}{gradient descent}
\acro{GI}{global illumination}
\acro{HVP}{Hessian-vector product}
\acrodefplural{HVP}{Hessian-vector products}
\acro{MC}{Monte Carlo}
\acro{MSE}{mean squared error}
\acro{NN}{neural network}
\acro{PDF}{probability density function}
\acro{RE}{rendering equation}
\acro{spp}{samples per pixel}
\acro{SPSA}{simultaneous perturbation stochastic approximation}
\end{acronym}

\newcommand{\task}[1]{\textsc{#1}}
\newcommand{\method}[1]{{\texttt{#1}}}
\newcommand{\mypara}[1]{\noindent\textbf{#1:}\quad}

\newcommand{\bftab}[1]{{\fontseries{b}\selectfont#1}}
\renewcommand{\eg}{\textit{e.g.}, }
\renewcommand{\ie}{\textit{i.e.}, }
\renewcommand{\wrt}{w.r.t.\ }

\newcommand{\hahvp}{Hessians and Hessian-vector products\xspace}

\def\paperID{8748} 
\def\confName{ICCV}
\def\confYear{2025}

\title{Stochastic Gradient Estimation for Higher-order Differentiable Rendering}
\author{Zican Wang\\
University College London\\
{\tt\small robert.wang.19@ucl.ac.uk}
\and
Michael Fischer\\
Adobe Research\\
{\tt\small mifischer@adobe.com}
\and
Tobias Ritschel\\
University College London\\
{\tt\small t.ritschel@ucl.ac.uk}
}

\maketitle

\begin{abstract}
We derive methods to compute higher-order differentials (\hahvp) of the rendering operator.
Our approach is based on importance sampling of a convolution that represents the differentials of rendering parameters and shows to be applicable to both rasterization and path tracing.
We further suggest an aggregate sampling strategy to importance-sample multiple dimensions of one convolution kernel simultaneously.
We demonstrate that this information improves convergence when used in higher-order optimizers such as Newton or Conjugate Gradient relative to a gradient descent baseline in several inverse rendering tasks. 
\end{abstract}

\mysection{Introduction}{introduction}

% what is differentiable rendering, why is it important 
Inverse rendering is concerned with optimizing the parameters of a scene to minimize a loss.
This can be a useful tool when the true parameters of the scene are unknown or non-trivial to set for a human expert, and therefore need to be inferred from observations or measurements. 
Examples for such scenarios are multi-view reconstruction \cite{mildenhall2021nerf} or the recovery of illumination and reflectance properties \cite{yu1999inverse}. 
Differentiable rendering has recently become a popular tool for this optimization, as \ac{AD} frameworks have become more widespread. 

% why is differentiable rendering hard? 
However, differentiating the rendering process is far from trivial, as the rendering function \cite{kajiya1986rendering} has zero and/or undefined gradients and step edges. 
Moreover, as many rendering operations rely on integration, discontinuities in the rendering pipeline cause problems for \ac{AD}-engines, as we can no longer na\"ively exchange the integral- and derivative-computations. 
% what have people done so far in inverse rendering, why rely on first order? 
Recent research therefore has designed a plethora of \emph{differentiable} renderers \cite{loper2014opendr}, which compute gradients in various ways, typically incurring more \cite{nimier2019mitsuba,liu2019soft} or less \cite{fischer2023plateau, fischer2024zerograds,deliot2024transforming} implementation and compute effort. 

\myfigure{Teaser}{Our approach allows sampling the Hessian for inverse rendering, here for the task of rotating the cup around its horizontal $(x_0)$ and vertical axes $(x_1)$. The estimated positive and negative gradients are shown in blue and red, respectively.}

% what are the benefits of higher order, why do we need it at all? 
A surprising insight is that, while these special renderers allow deriving gradients w.r.t. the scene parameters by differentiating (and back-propagating through) the rendering operator, virtually all of them are limited to first-order derivatives, and hence to expressing the gradient solely locally at a point in the parameter space.
At the time of writing, no attempts to derive the Hessians required for higher-order inverse rendering methods have been published.
However, decades of optimization research have shown the potential of higher-order methods in convergence and robustness \cite{nocedal1999numerical}.

% how does our method enable higher order, what does this mean? 
In this paper, we argue that these benefits are also applicable to inverse rendering scenarios and show that they translate to net-gains in optimization time and performance in differentiable rendering. 
We tackle the Achilles' heel of higher-order optimizers -- their increased per-iteration cost and computational / storage requirements -- by developing efficient \ac{MC} estimators of the required higher-order quantities that can be importance-sampled with established techniques \cite{owen2000safe}. 

% how do our results prove these claims / show these benefits 
Our method leads to speedups of 2.71$\times$ over previous methods, and net-gains in optimization time, speed and robustness while only assuming the rendering operator to be a black box that can be point-sampled. 

\mymath{\bandwidth}{\sigma}
\mymath{\brdf}{f_\mathrm r}
\mymath{\cdf}{P}
\mymath{\currentPoint}{\point^\step}
\mymath{\differentialOperator}{\operatorname D}
\mymath{\dimension}{n}
\mymath{\direction}{\mathbf v}
\mymath{\gradient}{g}
\mymath{\gradientOperator}{\operatorname D^\mathrm{G}}
\mymath{\gradientPDF}{\pdf^\mathrm G}
\mymath{\hessian}{\mathsf H}
\mymath{\hvp}{\hessian\gradient}
\mymath{\incoming}{\omega_\mathrm{i}}
\mymath{\kernel}{\kappa}
\mymath{\limval}{\varepsilon}
\mymath{\location}{\mathbf x}
\mymath{\hessianOperator}{\operatorname D^\mathrm{H}}
\mymath{\hvpOperator}{\operatorname D^\mathrm{HVP}}
\mymath{\renderingIntegrad}{R}
\mymath{\normal}{\mathcal N}
\mymath{\numsamples}{\mathrm{M}}
\mymath{\nextPoint}{\point^{\step+1}}
\mymath{\objective}{}
\mymath{\offset}{\tau}
\mymath{\offsets}{\bm\offset}
\mymath{\otherLocation}{\mathbf y}
\mymath{\outgoing}{\omega_\mathrm{o}}
\mymath{\pathSpace}{\Omega}
\mymath{\parameterspace}{\Theta}
\mymath{\parameters}{\bm\theta}
\mymath{\parameter}{\theta}
\mymath{\perturbation}{\alpha}
\mymath{\pdf}{p}
\mymath{\point}{\parameters}
\mymath{\radiance}{L}
\mymath{\random}{\xi}
\mymath{\residual}{\mathbf{r}}
\mymath{\smoothGradientOperator}{\operatorname{SD}^\mathrm{G}}
\mymath{\smoothHessianOperator}{\operatorname{SD}^\mathrm{H}}
\mymath{\smoothHVPOperator}{\operatorname{SD}^\mathrm{HVP}}
\mymath{\smoothingOperator}{\operatorname S}
\mymath{\smoothRadiance}{\bar{\radiance}}
\mymath{\step}{t}
\mymath{\stepSize}{\gamma}

% \begin{figure*}
%     \centering
%     \includegraphics[width=\linewidth]{Supplemental_images/Teaser.pdf}
%     \caption{}
%     \label{fig:Suzanne}
% \end{figure*}

\mysection{Previous Work}{previous_work}

\textbf{Gradient-based Optimization}, the main workhorse of modern neural network training and inverse rendering, uses the gradient of the objective function to take iterative steps in the parameter space towards an improved solution until a (local) minimum is found. 
The exact nature of these steps varies with the optimizer that is being used \cite{ruder2016overview}. 

First-order optimizers are simple to implement and cheap to execute but disregard important higher-order information about the shape of the objective function that could aid optimization, such as the second-order derivative. 
This usually comes at the expense of higher iteration counts, as many small steps are needed to converge to the solution. 
In contrast, higher-order optimizers incorporate information about the shape of the objective function, which usually allows them to take bigger steps in parameter space, leading to fewer iterations until convergence. 

For the specific case of \emph{second-order} optimization, this additional information is often provided via the Hessian \hessian (or approximations thereof) and the \ac{HVP}. 
The Hessian contains the second-order derivatives of the objective function \wrt the optimization parameters, and can be interpreted as the curvature of the objective function. 
As such, it can be used to inform the optimizer about how quickly the current gradient is changing and thus, in turn, about how large the optimization step should be. 

\textbf{Notation} The following will use lowercase boldface to denote vectors and uppercase sans to denote matrices, respectively.
\emph{Operators},  formally defined as functions acting on functions, will be denoted in uppercase Roman lettering in order to avoid confusion with regular functions. 

\paragraph{Gradient descent}
should be familiar to most readers, so we here recall only its terminology:
The first-order Taylor expansion of the cost function $f$ at position \point is
\begin{equation*}
    f(\point) \approx f(\currentPoint) + \gradient(\currentPoint)^\mathsf T (\point -\currentPoint),
\end{equation*}
where \gradient is the gradient $\nabla f=\mathrm df/\mathrm d \point$ of $f$.
The minimum is where the derivative is zero, which we can solve for as
\begin{equation}
   \frac{\textrm d}{\textrm d\point} f(\point) 
   =
   0
   \approx
   \gradient(\currentPoint)
   .
\end{equation}
As $f$ is a linear function only locally, we only make small steps with step size \stepSize, by an update direction $-\gradient(\currentPoint)$, as in 
\begin{equation*}
    \nextPoint = \currentPoint - \stepSize \gradient(\currentPoint).
\end{equation*}

\paragraph{Newton's method} is one of the most-used second-order optimizers. 
This term is derived from the second-order Taylor expansion of the objective $f$ around a point $\point$:
\begin{equation*}
    f(\point) \approx f(\currentPoint) + \gradient(\currentPoint)^\mathsf T (\point -\currentPoint) + \frac12
    (\point-\currentPoint)^\mathsf T
    \hessian(\currentPoint) 
    (\point-\currentPoint),
\end{equation*}
where \hessian is $\nabla^2 f = \mathrm d^2f/\mathrm d^2 \point$, the Hessian of $f$.

Ideally, we would like our update step to take us to an optimum.
There, the derivative is necessarily zero:
\begin{equation}\label{eq:newton}
    \frac{\textrm d}{\textrm d\point} f(\point) = 0 \approx \gradient(\currentPoint) + \hessian(\currentPoint) (\point - \currentPoint).
\end{equation}
Solving for \point then yields the update rule
\begin{equation}
   \nextPoint = \currentPoint -
    \stepSize
    \hessian^{-1}(\currentPoint)\gradient(\currentPoint), 
\end{equation}
where $\direction = - \hessian^{-1}(\currentPoint) \gradient(\currentPoint)$ is called the \emph{Newton direction}. 

Newton's method requires the computation, storage, and inversion of the full Hessian, which quickly becomes a bottleneck in higher dimensions, as the Hessian for an $n$-dimensional optimization problem is in $\mathbb{R}^{n\times n}$.

\paragraph{Newton Conjugate Gradient} \cite{shewchuk1994introduction}, upgrades Newton's method in two ways.
First, it solves for Newton's direction iteratively as per the linear equation $\hessian(\point) \direction = - \gradient(\point)$ using conjugate directions \cite{shewchuk1994introduction}.
Second, instead of an arbitrary step length $\stepSize$, it also decides the scalar \perturbation by which we move along this direction \direction.
To derive \perturbation, first consider the Taylor expansion:
\begin{align}
    \frac{\textrm d}{\textrm d \perturbation} f(\point + \perturbation \direction)
    \approx
    \gradient(\point)^\mathsf T \direction + \perturbation \direction^\mathsf T \hessian(\point) \direction &= 0
\end{align}
which can be re-arranged to
\begin{align}\label{eq:pertubation}
    \perturbation
    =
    -\frac
    {\gradient^\mathsf T(\point)\direction}
    {\direction^\mathsf T \hessian(\point) \direction}
    .
\end{align}
The update rule for the Newton Conjugate Gradient is more involved:
First, we maintain the direction \direction and a residual \residual, which are initialized to be the gradient at the initial positions:
\begin{equation}
    \direction^0 = 
    \residual^0 = 
    - \gradient(\currentPoint)
\end{equation}
We then find the  \perturbation by equation \refEq{pertubation} and update the point:
\begin{equation}
    \nextPoint = 
    \currentPoint + \perturbation \direction^\step
    .
\end{equation}
The next direction is chosen by  updating the residual \residual and computing a new conjugate search direction \direction with the Fletcher-Reeves formula~\cite{fletcher1963rapidly}:
\begin{align*}
    %\text{Direction update: }
    \residual^{\step+1} &= 
    \residual^\step - \perturbation \hessian(\point^\step) \direction^\step
    \\
    \beta &=
    \frac
    {\residual^{\step+1,\mathsf T} \residual^{\step+1}}
    {\residual^{\step,\mathsf T} \residual^\step}
    \\
    \direction^{\step+1} &=
    \residual^{\step+1} + \beta \direction^\step
    .
\end{align*}

%\paragraph{Positive definiteness}
Since Newton's method is derived via second-order Taylor expansion (approximated by a hyper-parabola), the derived Hessian may lead to either a maximum or a minimum depending on the local curvature of the function being approximated. 
This is determined by the positive-definiteness of the Hessian, and a negative-definite Hessian will diverge from the gradient's descent direction. 
To avoid this, one can ensure that the Hessian is always positive semi-definite using the Hessian modification technique and ensure step truncation to a trust region \cite{nocedal1999numerical}, $\alpha = \min( \alpha, \Delta \|\mathbf{v}\|^{-1})$,
where \(\Delta\) is the trust region radius (see Supplemental \refSec{PSD}). 

\paragraph{Hessian-vector product approximations} go one step further by entirely avoiding to compute \hessian when producing $\hessian\direction$.
\citet{pearlmutter1994fast} and \citet{werbos1988backpropagation} discuss different options to do so, but a simple option is central differences
\begin{equation}
    \hessian(\point) \direction \approx \lim _{\limval \rightarrow 0} \frac{\gradient(\point+ \limval  \direction)-\gradient(\point- \limval  \direction)}{2\limval}.
\end{equation}

%With these approximations, we are now equipped with the theoretical foundation to compute efficient approximations of the Hessians and \acp{HVP} required for our higher-order optimization scenarios. 
%We will briefly review existing work on higher-order optimization in the vision and graphics community in the following section before moving on to define our problem and approach in \refSec{our_approach}.

\paragraph{Hessians in vision and graphics} have been applied to several optimization tasks, for instance, for total-variation denoising
\cite{chan1999nonlinear},
texture parameterization and surface mappings with approximated Hessians \cite{sander2001texture}, 3D shape manipulation \cite{kilian2007geometric}, and image deconvolution tasks via (Hessian-free) Newton methods \cite{krishnan2009fast}.
For optical flow estimation, both \citet{zach2007duality} and \citet{werlberger2009anisotropic} use second-order information. 
Additionally, several works have explored Hessian-based algorithms in machine learning for vision tasks. 
\citet{yao2021adahessian} introduces a Hessian-based pruning method for CNNs for image classification, while \citet{ramesh2022hierarchical} utilizes Hessians to improve the training of image generation models and \citet{desai2022acorns} introduces an algorithm that derives Hessians for C code. 

Hessians in inverse rendering,
however, have received surprisingly little attention, potentially due to their additional implementation overhead and computational complexity.
In addition to the fact that conventional \ac{AD} systems are mostly designed for first-order gradient computations, calculating the second-order information requires storing the whole forward- and first-derivative--graph in memory, which can lead to exponential memory growth. 
\citet{nicolet2021large} are the closest to our work by approximating second-order steps for mesh reconstruction. 
However, they set the Hessian to the identity to avoid computational expense and instead work with a Laplacian regularizer, which works for their formulation and the case of mesh optimization, but is unclear how to translate to general problems. 

%TR: This is nice, butmaybe too far out
%For other derivative algorithms, \eg the adjoint formulation of light transport \cite{nimier2020radiative, vicini2021path}, it is unclear how one would efficiently derive the required Hessians, as the immediate approach -- differentiating the estimator of the adjoint equation -- would re-introduce many of the problems that the adjoint formulation was meant to solve in the first place \cite{nimier2019mitsuba}.

For derivative-free gradient estimators, it is equally unclear how second-order information would be computed. 
The zeroth-order estimators \ac{SPSA} \cite{spall1992multivariate} and \ac{FD} estimate a first-order gradient, whose second-order derivative naturally is zero. Extending these estimators to second-order information requires prohibitive amounts of function evaluations. 
ZeroGrads \cite{fischer2024zerograds}, which learns a neural network that fits the cost landscape, uses ReLU non-linearities, whose second-order derivative equally decays to zero. 
Covariance adaptation evolution strategy (CMA-ES) is often used when typical second-order optimization fails to converge \cite{hansen1996adapting}. 
It ensures that the Hessian is always positive definite, but if second-order derivative-based methods are successful, they are usually faster than CMA-ES.

%In this work, we make progress towards democratizing Hessian-based optimization in the inverse rendering community by efficiently computing the required higher-order quantities and by extending derivative-free approaches to the regime of higher-order optimization. 
%We hope that this article can bring this mindset to the inverse rendering community. 

\mysection{Our approach}{our_approach}

\mycfigure{Concept}{Comparison of classic (yellow) and smooth (red) gradients, as well as our Hessians (blue) on an inverse rendering problem to change the initial parameter \parameters so that the left triangle overlaps the right one.
Classic gradients are zero almost everywhere (plateaus) except where the triangles already overlap.
These methods do not converge.
Smooth gradients point into the right direction, but make steps far from the optimum (dotted line).
An update taking into account the curvature of the loss landscape lands at a point very close to the target. 
We incorporate this curvature information via our Hessians.}

We first describe the computation of gradients using importance sampling of a combined gradient-smoothing operator from previous work \cite{fischer2023plateau, chaudhuri2010smooth} which we then extend to Hessians, and in a next step to Hessian-vector products.

\mysubsection{Background}{background}

\paragraph{Rendering equation}
The \ac{RE} \cite{kajiya1986rendering} describes the radiance \radiance leaving a point \location in the scene into the direction \outgoing as 
\begin{equation}
\radiance(\location,\outgoing;\parameters)=
\int_\pathSpace
\underbrace{
\brdf(\incoming, \outgoing)
\radiance(\otherLocation, \incoming;\parameters)
}_{\renderingIntegrad(\incoming;\parameters)}
\mathrm d\incoming
,
\end{equation}
where \parameters are the parameters of the scene we would like to optimize, such as object geometry, reflectance, or light emission.
This integral over \pathSpace, \ie all incoming \incoming directions that multiply the radiance field arriving from that direction from the closest other point \otherLocation in direction \incoming with the \ac{BRDF} \brdf, has no analytical closed-form solution, and hence usually is approximated -- both in forward and inverse rendering -- via \ac{MC} methods.
We will shorthand the entire integrand as \renderingIntegrad. 

\paragraph{Problem statement}
We would now like to apply a differential operator \differentialOperator to the \ac{RE}, as in 
\begin{equation}
\differentialOperator\radiance(\location,\incoming;\parameters).
\end{equation}
If \differentialOperator was the gradient operator $\partial\radiance/\partial\parameters$, this would be differentiable rendering, for other operators, this becomes higher-order differentiable rendering.

The trouble is that we cannot move \differentialOperator, be it gradient or higher-order, into the integral, as in many cases (\eg for \ac{BRDF} or spatial derivatives), the integrand is discontinuous in \parameters, so
\begin{equation}
\differentialOperator
\radiance(\location,\incoming;\parameters)
\neq
\int_\pathSpace
\operatorname D
\renderingIntegrad(\incoming;\parameters)
\mathrm d\incoming
.
\end{equation}
However, the right-hand side of the above expression is exactly the quantity that na\"ively-applied \ac{AD} computes \cite{vicini2021path, nimier2020radiative}, leading to wrong gradients in (any-order) differentiable rendering. 

\paragraph{Solution}
The idea is to enforce the property that prevents differentiation -- smoothness --, so that we actually can differentiate.
To that end, assume a further linear operator \smoothingOperator that is smoothing any function in \parameters.
This provides a smooth rendering equation \smoothRadiance
\begin{equation}
\smoothRadiance(\location,\outgoing;\parameters)=
\int_\pathSpace
\smoothingOperator
\renderingIntegrad(\incoming;\parameters)
\mathrm d\incoming
.
\end{equation}

Smoothing can be achieved by convolution, so for any function $f$ in any dimension
\begin{equation}
    \smoothingOperator f(\parameters)=
    \int_\parameterspace
    \kernel(\offsets)
    f(\parameters-\offsets)
    \mathrm d\offsets
    ,
\end{equation}
where \kernel is a smoothing kernel, such as a Gaussian, which we use in this work.
This convolved integrand is now smooth, which, according to Leibniz' rule, allows us to move the differential operator into the integral
\begin{equation}
\differentialOperator \smoothRadiance(\location,\outgoing;\parameters) =
\int_\pathSpace
\differentialOperator
\smoothingOperator
\renderingIntegrad(\incoming;\parameters)
\mathrm d\incoming
,
\end{equation}

which, after rearranging and expanding, yields an integral that can be approximated via \ac{MC}:
\begin{equation}
\differentialOperator \smoothRadiance(\location,\outgoing;\parameters) =
\int_\pathSpace
\int_\parameterspace
\operatorname D
\kernel(\offsets)
\renderingIntegrad(\incoming;\parameters-\offsets)
\mathrm d\offsets
\mathrm d\incoming
.
\end{equation}

\ac{MC} here means to take random samples from the product space of light paths and scene parameters.
This works best if we can importance-sample for the integrand.
The integrand here is a product of four terms.
Sampling for the incoming radiance and \ac{BRDF} terms has been investigated in the rendering community \cite{veach1995optimally} and is not our consideration here, so we simply adopt these strategies. 
Sampling for the application of the differential operator to the smoothing kernel is the essence of the problem at hand.
Depending on the choice of differential operator, we will derive three sampling strategies for the three resulting estimators next. 

\paragraph{Conclusion}
In conclusion, to perform efficient and practical any-order differentiation of the \ac{RE}, we would need to implement two functions: first, a convolution kernel that combines smoothing and the desired differentiation, and second, a function to sample from that kernel for importance sampling.
We will now do so for the gradient (\refSec{gradient}), Hessian (\refSec{hessian}) and \ac{HVP} (\refSec{hessian_vector_product}).

\myfigure{Derivation}{Derivation of smooth differentiation by convolution for three differential operators (rows) involves three steps (columns): Defining the operator to combine smoothing and differentiation (1st col.), positivization and normalization to become a \ac{PDF} (2nd col.), creating an inverse mapping (3rd col.), which finally allows sampling (4th col.). }

\mysubsection{Gradients}{gradient}

\mypara{Operator}
For first-order gradient descent, \citet{fischer2023plateau} have differentiated using the gradient operator 
\begin{equation}
\gradientOperator
=
\nabla
=
\partial/\partial\offset_i
\in
(\mathbb R^n\rightarrow\mathbb R)
\rightarrow
(\mathbb R^n\rightarrow\mathbb R^n)
,
\end{equation}
which maps a scalar function of $n$ dimensions to an $n$-dimensional gradient vector field.

The combination of smoothing and differentiation is
\begin{equation}\label{eq:gradient}
\gradientOperator\kernel(\offsets)=
\nabla_i \normal(\offsets, \bandwidth)=
-\frac{\offset_i}{\bandwidth^2} \cdot \normal(\offsets, \bandwidth).
\end{equation}

For a derivation, please see the supplemental, \refSec{gradient_derivation}.

\mypara{Sampling}
For sampling a one-dimensional Gaussian gradient, we can use inverse transform sampling via the Smirnov transform \cite{fischer2023plateau}.
To this end, the integrand has to be a \ac{PDF}, \ie positive and integrating-to-1 \cite{owen2000safe}.
Subsequently, we compute the integral of a positivized version of $\pdf^\mathrm G$, the \ac{CDF} $\cdf^\mathrm G$, and invert it as
\begin{equation}
\cdf^{\mathrm G,-1}(\random) =
\begin{dcases}
-\sqrt{2\bandwidth^2\log(2\random)}&
\text{if\,}\random\leq 0.5,\\
+\sqrt{2\bandwidth^2\log(1-\random)
}&
\text{else}.
\end{dcases}
\end{equation}
This derivation applies to the dimension which is being differentiated. 
The separability of multi-dimensional Gaussians ensures that the other dimensions remain a Gaussian distribution, and we can sample these dimensions independently.
For a derivation of this, see Suppl. \refSec{gradient_sampling_derivation}.

\mysubsection{Hessians}{hessian}

\mypara{Operator}
The differential operator for Hessians is
\begin{equation}
\hessianOperator
=
\nabla^2
=
\partial^2/\partial\offset_i\partial\offset_j
\in
(\mathbb R^n\rightarrow\mathbb R)
\rightarrow
(\mathbb R^n\rightarrow\mathbb R^{n\times n})
,
\end{equation}
which maps a scalar function in $n$ dimensions to its $n\times n$-element Hessian field.
The combination of second-order derivatives and Gaussian smoothing is
\begin{equation}
\hessianOperator\kernel_{i,j}(\offsets)=
\begin{dcases}
\left(
-
\frac{1}{\bandwidth^2}
+
\frac{\offset_i^2}{\bandwidth^4}
\right)
\cdot \normal(\offsets, \bandwidth)&
\text{if }i=j,
\\
\frac{\offset_i\offset_j}{\bandwidth^4} \cdot \normal(\offsets, \bandwidth)&
\text{else.}
\end{dcases}
\end{equation}
For a derivation, see supplemental \refSec{hessian_derivation}.

\mypara{Sampling}
We first need to positivize as the function is signed, then scale the function so that it is a valid distribution.
This is done differently for diagonal and off-diagonal elements.
For the diagonal case, we construct the \ac{CDF} of a second-order derivative $\cdf^{\mathrm H}_{ii}$ of the 2D Gaussian:
\begin{equation}
\cdf^{\mathrm H}_{ii}(\offset_i = u)=
\begin{dcases}
    \hfill
    -\frac{u}{4\bandwidth}
    \exp\left(\frac{1}{2}-\frac{u^2}{2\bandwidth^2}\right)
    &
    \text{if }u<-\bandwidth,\\
    \hfill
    0.5+
    \frac{u}{4\bandwidth}
    \exp\left(\frac{1}{2}-\frac{u^2}{2\bandwidth^2}\right)&
    \text{if } 
    %-\bandwidth<\offsets<\bandwidth,
    u \in \left[-\bandwidth, \bandwidth \right]
    \\
    \hfill
    1-
    \frac{u}{4\bandwidth}
    \exp\left(\frac{1}{2}-\frac{u^2}{2\bandwidth^2}\right)&
    \text{if }u>\bandwidth.\\
\end{dcases}
\end{equation}
See supplemental \refSec{sampling_hessian} for a derivation of this result.
This is a transcendental equation
%, there is no analytical form of expressing the inverse \ac{CDF} \cite{boyd2014solving}.
whose inverse \ac{CDF} cannot be expressed in closed form via elementary functions \cite{boyd2014solving}.
Instead, we 1D-tabulate its values on the range $10\sigma$ for every $i$.
The inverse function is found by searching in that range, and storing the pre-sorted inverse indices for $O(1)$ access.

The off-diagonals are the product of two partial gradients that we have already derived in \refEq{gradient}.
Fortunately, the product of two independent distributions can be sampled by sampling each one independently. Thus, no extra derivation is required here.
For sampling in higher dimensions, similar to the gradient sampler in \refSec{gradient_sampling_derivation}, the rest of the dimensions are sampled from a Gaussian distribution.

In all methods, we exploit the symmetry in the Hessian matrix when sampling and estimating.

\mysubsection{Hessian-vector product}{hessian_vector_product}

\mypara{Operator}
As explained in \refSec{previous_work}, avoiding to store the full Hessian is possible by using \acp{HVP}:
\begin{equation}
\hvpOperator =
\nabla^2\direction
=
\partial^2/\partial\offset_i\partial\offsets
\cdot
\partial/\partial\offsets
.
\end{equation}
In essence, a \ac{HVP} is the directional gradient of the gradient.
As there is one direction and the gradient has $n$ dimensions, the \ac{HVP} is an $n$-dimensional vector, too.
In contrast, a Hessian is the non-directional gradient of the gradient, \ie the gradient along all $n$ dimensions, and as such in $\mathbb{R}^{n\times n}$. 
%As each gradient has $n$ dimensions and there is $n$ of them, the Hessian is an $n \times n$ matrix.

\mypara{Sampling}
To sample the smooth \ac{HVP}, we simply need the directional central differences of an estimator of gradients, which we already have.
So the estimator is the difference of two first-order estimators, evaluated at positions shifted from the current solution along the gradient direction \cite{pearlmutter1994fast,werbos1988backpropagation}:
\begin{equation}
\hvpOperator\kernel(\offsets)=
\frac{
\gradientOperator\kernel
(\offsets-\varepsilon\direction)
-
\gradientOperator\kernel
(\offsets+\varepsilon\direction) 
}{2\varepsilon}.
\end{equation}

\mysubsection{Aggregate}{Aggregate}

We summarize the key property of all samplers -- the number of function evaluations required -- in \refTab{Complexity}.
Function evaluations require execution of the black-box rendering engine, the most costly part of the optimization, and hence should be minimized. 
\begin{wraptable}{l}{42mm}
\small
    \centering
    \caption{Time complexity of different estimator variants. AIS: aggregated importance sampling.}
    \label{tab:Complexity}
\begin{tabular}{lccc}
\toprule
& \multicolumn1c{no IS}
& \multicolumn1c{IS} 
& \multicolumn1c{AIS} \\
\midrule
Fin. Diff. & $n$ & $n$ & $n$ \\
FR22 \cite{fischer2023plateau} & $1$ & $1$ & $1$ \\
\midrule
Our Grad. & $1$ & $n$ & $1$ \\
Our Hess. & $1$ & $n^2$ & $1$ \\
Our HVP & $1$ & $n$ & $1$ \\
\bottomrule
\end{tabular}
\end{wraptable}
Classic finite differences %(first row in \refTab{Complexity})
take opposing samples in each dimension and hence, for an $n$-dimensional problem, require $2n$ function evaluations for a single gradient sample.
% \begin{wraptable}{l}{42mm}
% \small
%     \centering
%     \caption{Estimator complexities.}
%     \label{tab:Complexity}
% \begin{tabular}{lccc}
% \toprule
% & \multicolumn1c{no IS}
% & \multicolumn1c{IS} 
% & \multicolumn1c{AIS} \\
% \midrule
% Fin. Diff. & $n$ & $n$ & $n$ \\
% FR22 \cite{fischer2023plateau} & $1$ & $1$ & $1$ \\
% \midrule
% Our Grad. & $1$ & $n$ & $1$ \\
% Our Hess. & $1$ & $n^2$ & $1$ \\
% Our HVP & $1$ & $n$ & $1$ \\
% \bottomrule
% \end{tabular}
% \end{wraptable}
All other methods (second row in \refTab{Complexity} onward) are based on \ac{MC}, so we can get a gradient estimate using a fixed number of \numsamples function evaluations.
This, however, comes at the cost of variance, which can be reduced with importance sampling, but that is done in all dimensions of the differential quantity independently, so it requires as many evaluations as these have elements (second column in \refTab{Complexity}).
For a Hessian, this can be substantial, $n^2$.

\begin{figure}
    \centering
    \includegraphics{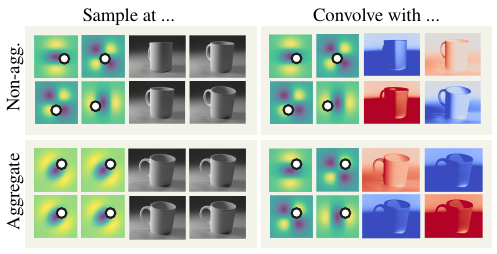}
    \vspace{-3mm}
    \caption{Aggregate and non-aggregate sampling for a 2D optimization space and, consequentially, a $2\times2$ Hessian:
Without aggregation, importance sampling is done for each element of the Hessian independently (four points at different positions in each kernel), leading to four rendering calls and four different mugs.
Each value is then weighted by the kernel (blue and red colors denote positive and negative).
In aggregate sampling, we importance-sample the average of all four stencils, resulting in a single sample location and, hence, four times the same mug that can be rendered in one call, weighted with four different kernels.}
    \label{fig:Aggregate}
    \vspace{-6mm}
\end{figure}
%\myfigure{Aggregate}{Aggregate and non-aggregate sampling for a 2D optimization space and, consequentially, a $2\times2$ Hessian:
%Without aggregation, importance sampling is done for each element of the Hessian independently (four points at different positions in each kernel), leading to four rendering calls and four different mugs.
%Each value is then weighted by the kernel (blue and red colors denote positive and negative).
%In aggregate sampling, we importance-sample the average of all four stencils, resulting in a single sample location and, hence, four times the same mug that can be rendered in one call, weighted with four different kernels.}

As a compromise between low number of function evaluations and low variance, we propose ``aggregate'' importance sampling of the differential quantity (\refFig{Aggregate}): instead of importance-sampling each dimension optimally but individually, we importance-sample \wrt the \emph{average} of the convolution across \emph{all} dimensions (combining several proposals by sampling from a mixture is similar to MIS using the balance heuristic \cite{veach1995optimally}). 
This average convolution is a single function again, and with our combined sampling strategy can again be sampled with one function call per iteration (last column of \refTab{Complexity}).
We implement this in two simple steps: first we randomly decide which element of the differential representation to choose (\eg which matrix element out of $n^2$ in the Hessian), and then we compute the convolution sample for all $n^2$ but with the same function value that is only evaluated once.
This increases variance, as the average kernel is not identical to the individual kernels but shares many properties, \eg they are all zero in value at position 0.
As a linear combination of unbiased estimators, this is still unbiased.
We show in \refSec{VarianceAnalysis} that this pays off.
Note that \refTab{Complexity} is both time and space complexity, except for the aggregate importance sampling of the \ac{HVP}, where the space complexity stays $O(n)$ while time complexity is reduced to $O(1)$.

\mysection{Evaluation}{evaluation}

\mysubsection{Methods}{methods}
Our evaluation compares the performance of gradients (input to gradient descent), Hessians and non-aggregate as well as aggregate Hessian-vector product (both input to higher-order optimizers). 

\paragraph{Baselines}
An established inverse rendering solution is 
\method{Mitsuba} with gradient descent.
\method{FR22} is using gradient descent with \citet{fischer2023plateau}  MC gradients.
\method{OursG}, \method{OursH},
\method{OursHVP} and \method{OursHVPA} are our approach for gradient, \hahvp, and aggregates, respectively.
\method{OursG} is used with \ac{GD}, the others in combination with \ac{CG}.
We have also experimented with \method{LBFGS} as detailed in the supplemental, but do not report it here as it almost never converges and never at competitive speed.
All methods use the Adam optimizer.

\paragraph{Tasks}
We tackle artificial problems with known analytic \hahvp as well as real inverse rendering tasks.
As a simple first test case, we optimize the smooth, quadratic potential(\task{Quad}\refFact{quad}) function $ax_{0}^{2} + bx_{1}^{2} + cx_{0}x_{1}$ in $\mathbb{R}^2$, with the fixed variables $a=5, b=5, c=7.5$.

Second, we optimize the classic plateau-demonstration task from \cite{fischer2023plateau}: the 2D position of some boxes is optimized to match a reference. 
This is already a much harder task, as there is a plateau in the cost landscape when the squares do not overlap (almost always the case in the initial configuration), as the image-space error does not change. 
We study the case of one (\task{Box2}\refFact{box2}) and five (\task{Box10}\refFact{box10}) squares, with two and ten dimensions to optimize, respectively.

\newcommand{\tabhead}[1]{\multicolumn1c{#1}}

\newcolumntype{M}{>{\collectcell\method\scriptsize}{l}<{\endcollectcell}}
\newcolumntype{T}{>{\collectcell\small}{r}<{\endcollectcell}}

\newcommand{\na}{\multicolumn1c{---}}
\newcommand{\resultSplit}{\cmidrule(lr){1-9}}

\newcommand{\iconcell}[2]{\multirow{5}{*}{\makecell{\task{#1}\fact{#2}\\\includegraphics[width=1.5cm]{images/task/#1}}}}

\begin{table*}[]
    \renewcommand{\tabcolsep}{0.2cm}
    \centering
    \caption{
    Quantitative results of different methods on different tasks (rows) and their convergence plots.
    We report convergence time in wall-clock units, in ratio to the overall best method, \method{OurHVPA}.
    In the numerical columns, $.9$ and $.99$ report the time taken to achieve $90$ and $99$\% error reduction from the initial starting configuration, respectively, while the bar plots graphically show these findings.
    The line plots report image- and parameter-space convergence in the left and right column, respectively, on a log-log scale.}
    \label{tab:Results}    
    %\footnotesize
    \scriptsize
    \begin{tabular}{c M rrr rrr c}
    \toprule
    \tabhead{Task}&   
    \tabhead{Method}&   
    \multicolumn3c{Image error}&
    \multicolumn3c{Parameter error}&
    %\multirow{1}{*}{\includegraphics[width=8.8cm]{images/Table}}
    \multirow{1}{*}{\includegraphics{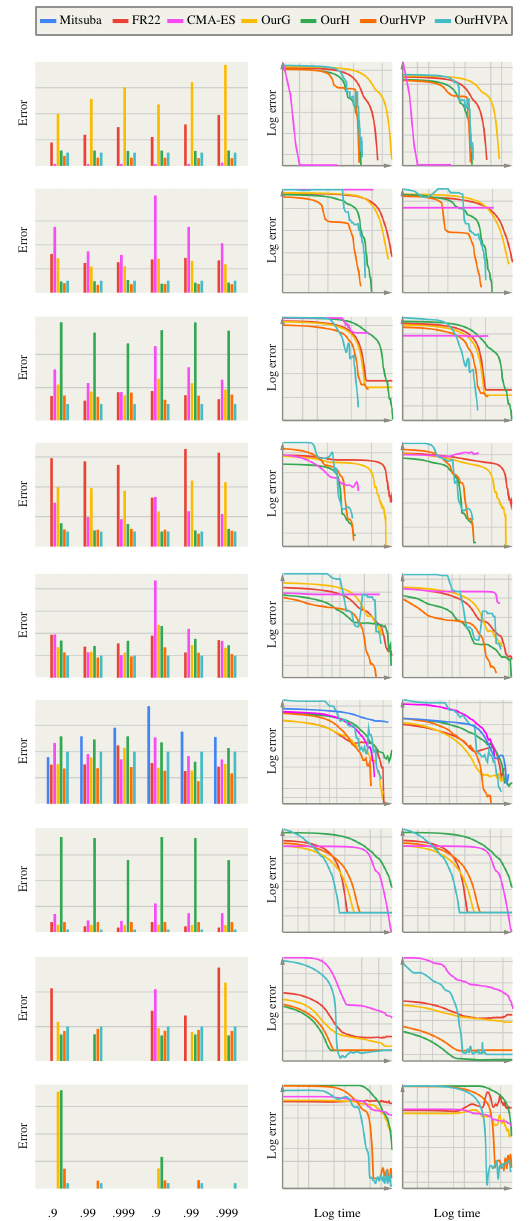}}
    \\
    \cmidrule(lr){3-5}
    \cmidrule(lr){6-8}
    &&
    \tabhead{\scriptsize.9}&
    \tabhead{\scriptsize.99}&
    \tabhead{\scriptsize.999}&
    \tabhead{\scriptsize.9}&
    \tabhead{\scriptsize.99}&
    \tabhead{\scriptsize.999}
    \\
    \midrule
\iconcell{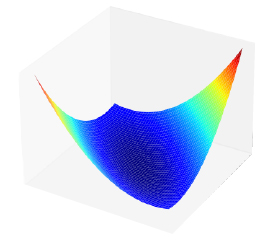}{quad} & Mitsu & \na\fact{WorksForAllFunc} & \na & \na & \na & \na & \na \\
& FR22 & 1.80 & 2.38 & 2.98 & 2.20 & 3.17 & 3.90 \\
& CMA-ES & \bftab{0.15} & \bftab{0.13} & \bftab{0.14} & \bftab{0.13} & \bftab{0.14} & \bftab{0.23} \\
& OurG & 3.98 & 5.13 & 6.02 & 4.72 & 6.42 & 7.77 \\
& OurH & 1.17 & 1.17 & 1.15 & 1.18 & 1.16 & 1.17 \\
& OurHVP & \underline{0.76} & \underline{0.62} & \underline{0.62}  & \underline{0.63}  & \underline{0.59}  & \underline{0.60}  \\
& OurHVPA & 1.00 & 1.00 & 1.00 & 1.00 & 1.00 & 1.00 \\
\resultSplit
\iconcell{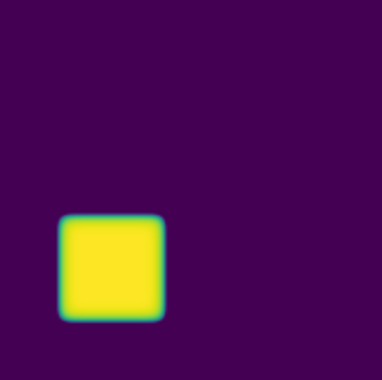}{box2} & Mitsu & \na & \na & \na & \na & \na & \na \\
& FR22 & 3.24 & 2.49 & 2.57 & 2.78 & 2.89 & 2.70 \\
& CMA-ES & 5.51 & 3.45 & 3.15 & 8.13 & 5.51 & 4.13 \\
& OurG & 2.88 & 2.16 & 2.21 & 2.86 & 2.65 & 2.38 \\
& OurH & \underline{0.93} & \underline{0.95} & 1.08 & \underline{0.76} & \underline{0.84} & 1.84 \\
& OurHVP & \bftab{0.81}  & \bftab{0.64}  & \bftab{0.70} \fact{HVPFasterForLowDimension} & \bftab{0.74}  & \bftab{0.72}  & \bftab{0.72}  \\
& OurHVPA & 1.00 & 1.00 & \underline{1.00} & 1.00 & 1.00 & \underline{1.00} \\
\resultSplit
\iconcell{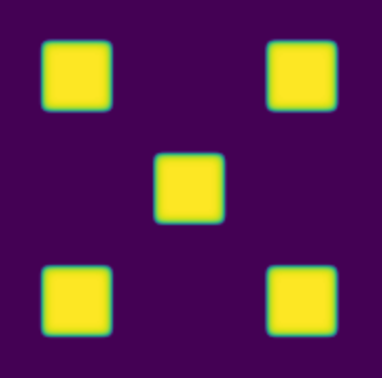}{box10} & Mitsu & \na & \na & \na & \na & \na & \na \\
& FR22 & \underline{1.49} & \underline{1.20} & 1.71 & 1.80 & 1.54 & \underline{1.29} \\
& CMA-ES & 3.10 & 2.27 & 1.73 & 4.52 & 3.23 & 2.48 \\
& OurG & 2.18 & 1.77 & \underline{1.53} & 2.53 & 2.26 & 1.88 \\
& OurH & 5.97 & 5.33 & 4.68 & 5.48 & 5.96 & 5.45 \\
& OurHVP & 1.51 & 1.44 & 1.69\fact{HVPSlowerForHighDimension} & \underline{1.26} & \underline{1.50} & 1.58 \\
& OurHVPA & \bftab{1.00}  & \bftab{1.00}  & \bftab{1.00}  & \bftab{1.00}  & \bftab{1.00}  & \bftab{1.00}  \\
\resultSplit
\iconcell{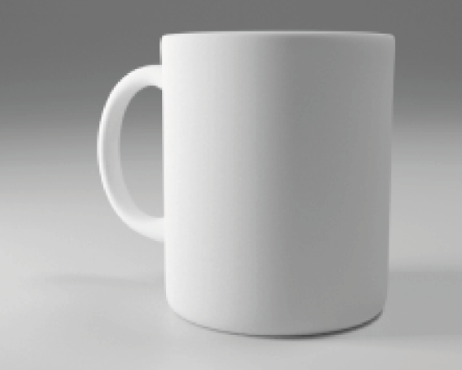}{mug} & Mitsu & \na & \na & \na & \na & \na & \na \\
& FR22 & 5.92 & 5.69 & 5.46 & 3.28 & 6.53 & 6.29 \\
& CMA-ES & 2.95 & 1.97 & 1.82 & 3.32 & 2.38 & 2.20 \\
& OurG & 3.99 & 3.91 & 3.74 & 2.36 & 4.42 & 4.30 \\
& OurH & 1.57 & \underline{1.09} & 1.53 & \underline{1.00} & 1.08 & 1.20 \\
& OurHVP & \underline{1.15} & 1.14 & \underline{1.18} & 1.14 & \bftab{0.87}  & \underline{1.05} \\
& OurHVPA\fact{HVPA1DIM} & \bftab{1.00}  & \bftab{1.00}  & \bftab{1.00}  & \bftab{1.00}  & \underline{1.00} & \bftab{1.00}  \\
\resultSplit
\iconcell{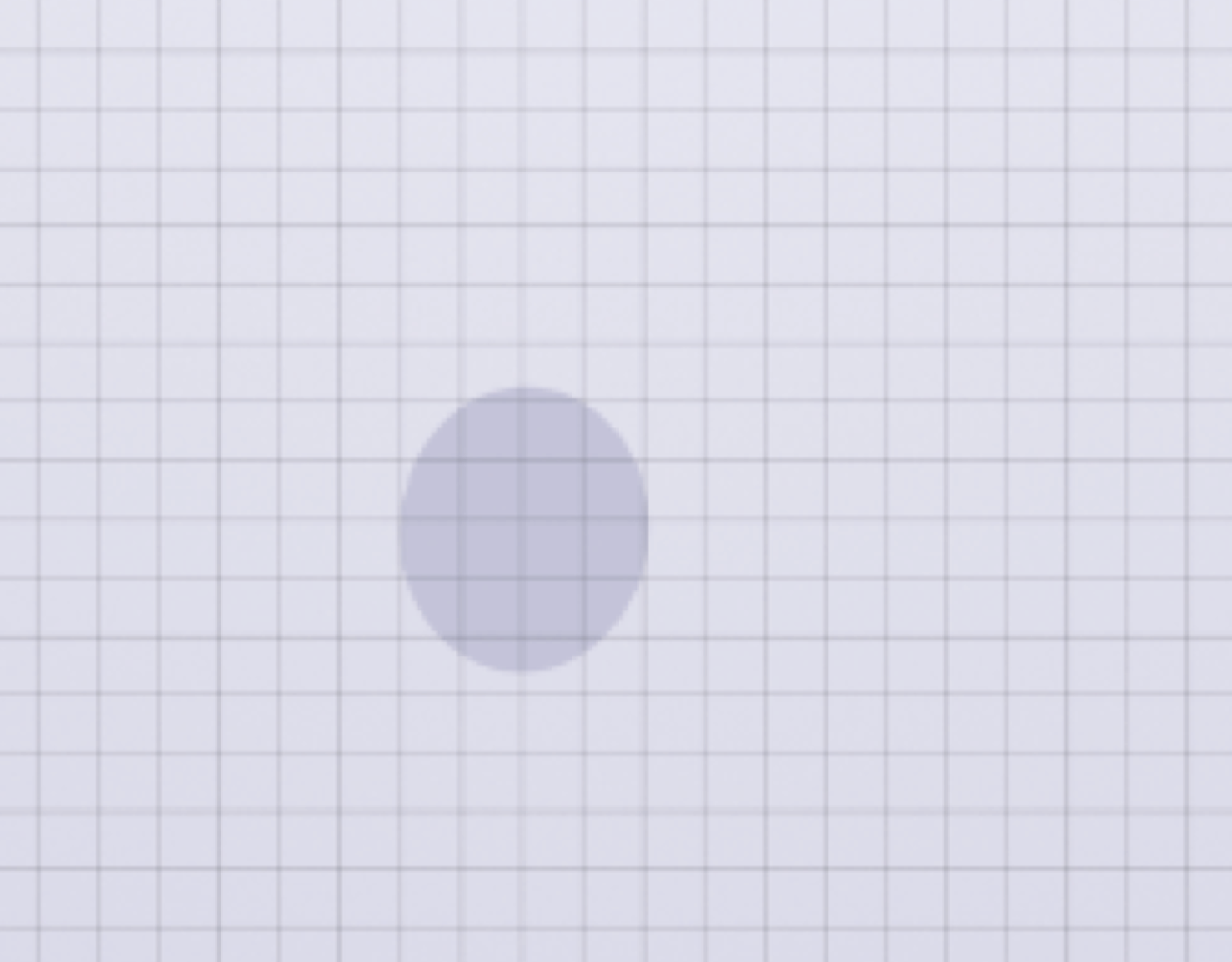}{shad} & Mitsu & \na & \na & \na & \na & \na & \na \\
& FR22 & 1.94 & 1.40 & 1.55 & 1.89 & 1.14 & 1.71 \\
& CMA-ES & 1.96 & 1.15 & 1.03 & 4.38 & 2.20 & 1.66 \\
& OurG & 1.37 & 1.18 & 1.13 & 2.38 & 1.47 & 1.34 \\
& OurH & 1.67 & 1.44 & 1.66 & 2.32 & 1.74 & 1.46 \\
& OurHVP & \underline{1.14} & \bftab{0.91}  & \bftab{0.96}  & \underline{1.37} & \underline{1.13} & \underline{1.07} \\
& OurHVPA & \bftab{1.00}  & \underline{1.00} & \underline{1.00} & \bftab{1.00}  & \bftab{1.00}  & \bftab{1.00}\fact{HVPABestForShad} \\
\resultSplit
\iconcell{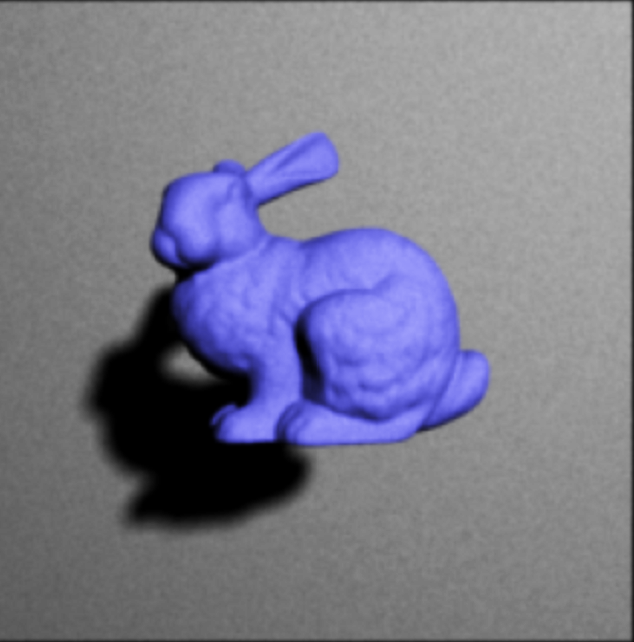}{bunny} & Mitsu\fact{MitsubaConvergesOnBunny} & 0.90 & 1.29 & 1.46 & 1.87 & 1.38 & 1.28 \\
& FR22 & \underline{0.75} & \underline{0.75} & 1.12 & 0.78 & \underline{0.63} & \underline{0.71} \\
& CMA-ES & 1.16 & 0.95 & 0.85 & 1.27 & 0.91 & 0.85 \\
& OurG & 0.76 & 0.89 & \underline{1.07} & \underline{0.69} & 0.64 & 0.76 \\
& OurH & 1.29 & 1.23 & 1.29 & 1.18 & 0.80 & 1.07 \\
& OurHVP & \bftab{0.68}  & \bftab{0.68}  & \bftab{0.70}  & \bftab{0.63}  & \bftab{0.44}  & \bftab{0.58}  \\
& OurHVPA & 1.00 & 1.00 & 1.00 & 1.00 & 1.00 & 1.00 \\
\resultSplit
\iconcell{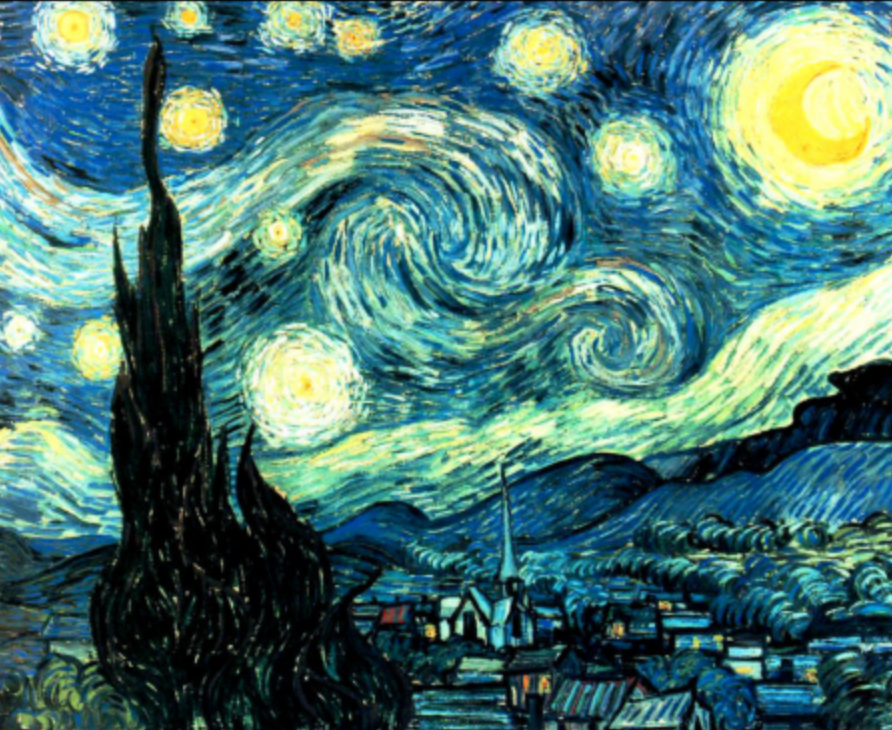}{texture} & Mitsu & \na & \na & \na & \na & \na & \na \\
& FR22 & 4.04 & 2.27 & 1.80 & 4.04 & \underline{2.27} & \underline{1.80} \\
& CMA-ES & 7.08 & 4.60 & 4.35 & 11.16 & 7.47 & 7.44 \\
& OurG & \underline{3.03} & \underline{2.86} & \underline{2.82} & \underline{3.03} & 2.86 & 2.82 \\
& OurH\fact{HBadInHighDimensions} & 40.12 & 39.78 & \na & 40.12 & 39.78 & \na \\
& OurHVP & 4.01 & 4.00 & 4.00 & 4.01 & 4.00 & 4.00 \\
& OurHVPA & \bftab{1.00}  & \bftab{1.00}  & \bftab{1.00}  & \bftab{1.00}  & \bftab{1.00}  & \bftab{1.00}  \\
\resultSplit
\iconcell{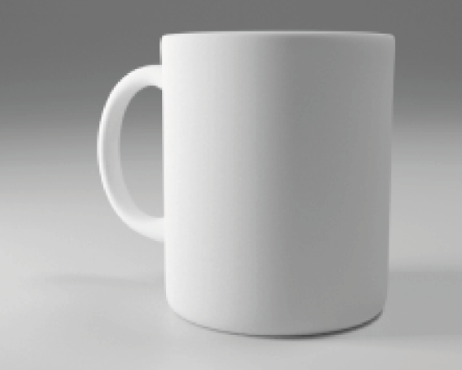}{cnn} & Mitsu & \na & \na & \na & \na & \na & \na \\
& FR22 & 2.11\fact{FR22SlowerOnCNN} & \na & \na & 1.45 & 1.32 & 2.71 \\
& CMA-ES & \na & \na & \na & 2.08 & \na & \na \\
& OurG & 1.14 & \na & \na & 0.96 & 0.83 & 2.28 \\
& OurH & \bftab{0.77} & \bftab{0.77} & \na & \bftab{0.74} & \bftab{0.77} & \bftab{0.74} \\
& OurHVP & \underline{0.87} & \underline{0.93}\fact{OurHVPNotFasterOnCNN} & \na & \underline{0.88} & \underline{0.91} & \underline{0.87} \\
& OurHVPA & 1.00 & 1.00 & \na\fact{OurHVPANotOnCNN} & 1.00 & 1.00 & 1.00 \\
\resultSplit
\iconcell{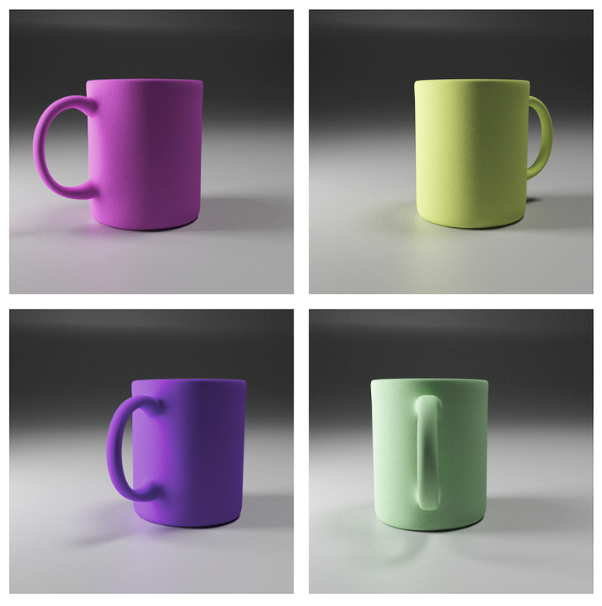}{cnn5} & Mitsu & \na & \na & \na & \na & \na & \na \\
& FR22 & \na & \na & \na & \na & \na & \na \\
& CMA-ES & \na & \na & \na & \na & \na & \na \\
& OurG & 18.68 & \na & \na & 3.70 & \na & \na \\
& OurH & 19.00 & \na & \na & 5.79 & \na & \na \\
& OurHVP & \underline{3.64}\fact{OurHVPSlowerForCNN5} & \underline{1.46} & \na & \underline{1.49} & \underline{1.58} & \na \\
& OurHVPA & \bftab{1.00} & \bftab{1.00} & \na & \bftab{1.00} & \bftab{1.00} & \bftab{1.00} \\
    \bottomrule
    \\
    \end{tabular}
\end{table*}

For real inverse problems, we study the optimization of reflectance, light, and geometry and render using Mitsuba. 
In the \task{Mug}\refFact{mug} task, we optimize the vertical rotation of a coffee cup such that it matches a reference. 
In the \task{Shadow}\refFact{shad} task, we optimize the position of a sphere that is unobserved in the rendered image, such that the shadow it casts matches a reference shadow. 
In the \task{Bunny}\refFact{bunny} task, the $x$ and $z$ position and the rotation around the $z$ axis of the Stanford bunny are optimized.
This task is set up specifically for \method{Mitsuba} to converge\refFact{MitsubaConvergesOnBunny}, with no plateaus in the loss function.

\begin{figure*}
    \centering
    \includegraphics[width=\linewidth]{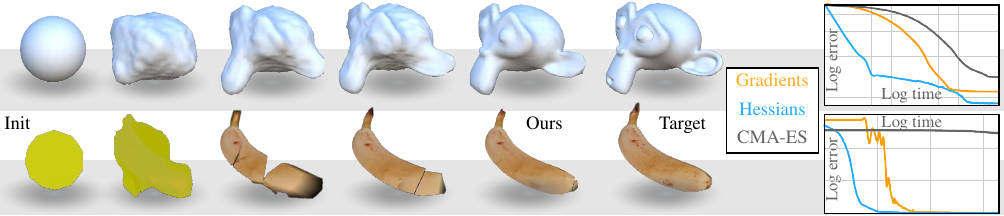}
    \vspace{-6mm}
    \caption{Inverse mesh optimization from renderings. Gradients use \method{FR22}, Hessians use \method{OursHVPA}, CMA-ES uses the pycma package.}
    \label{fig:meshoptim}
    \vspace{-5mm}
\end{figure*}

To test the scaling with dimensionality, we also optimize the pixels of a 32$\times$32 texture (\task{texture}\refFact{texture}), with \num{1024} parameters.
In addition, we have used our approach on a mesh optimization task, where we fit the 3D position of the vertices of a tessellated sphere to images of a target object. 
For \task{Suzanne}, we fit 2,562 vertices, for \task{Banana}, we fit the 2,737 vertices and their RGB vertex colors, similar to the DTU setup \cite{jensen2014large}. 
For trajectories of the optimization and the outcome, see \refFig{meshoptim} top and bottom rows, respectively. 

Finally, we learn a \task{CNN}\refFact{cnn} to predict scene parameters from images using inverse rendering without scene parameter labels.
In this case, the optimization is learning in the stricter sense: we tune parameters of a deep architecture instead of scene parameters directly.
The CNN takes as input an image of a mug and produces as output the orientation for a mug.
As a loss, these parameters are inserted into a renderer, differentiated by different approaches and compared to a target.
The CNN has \num{267745} parameters, learned by differentiating through an image loss and rendering of the scene given the estimated parameters. Another CNN (\task{CNN5}\refFact{cnn5}) is trained to predict the rotation, position, and color of the mug with \num{268773} parameters.

%\mycfigure{meshoptim}{Inverse mesh optimization from renderings. Gradients uses \method{FR22}, Hessians uses \method{OursHVPA}, CMA-ES uses the pycma package.}

Differentials of this function are computed as detailed in supplemental \refSec{combination}, where we derive a grey-box approach that samples only the black-box part (the rendering) and combines its differentials with analytic differentials for the white-box part (the CNN).
Importantly, this idea works for gradients, as well as for higher-order differentials. % such as the ones we study here.

\vspace{-3mm}
\paragraph{Metrics}
Our main measure of success is convergence speed in wall-clock and parameter difference to the true parameters (which we know in all inverse rendering tasks as we created the scenes).
A secondary measure of success is the image difference, as during optimization, we consider the parameters hidden.
All metrics are evaluated across an ensemble of 20 runs averaged across 10 steps in time.
In convergence plots, the ensemble median is shown at every point in time, averaged across 20 time steps.

\mysubsection{Results}{results}

The main results of our evaluation are summarize in \refTab{Results} where empty cells did not converge or the method is not applicable to that task.
On average across tasks and methods, our premiere method \method{oursHVPA} speeds up the convergence by a factor of 2.71.
%This ratio varies between different tasks. 

In general, our methods, in particular \method{OursHVP} and \method{OursHVPA}, lead the level of error reduction across all time budgets, as seen by comparing the convergence curves vertically where they consistently decrease fastest.

We benefit most in the artificial \task{Quad} task, but \task{Mug} and \task{Shadow} are both real rendering problems, where we are around seven times and 60\,\% faster, respectively.
We see that \method{OurG} typically cannot outperform \method{FR22}, which uses an approximation of the correct high-dimensional gradient kernel.
Doing it ``right'' only pays off when going to higher order.
We also note that \method{OursHVP} outperforms \method{OursH} due to less function evaluations (as the metric is wall-clock - the iteration count for both is close).
The log-log plots confirm that the higher-order variants enabled by our approach are much faster, but also converge slightly less stable.

While tasks like \task{quad} and \task{Box2} are not applicable to inverse renderers like \method{Mitsuba}\refFact{WorksForAllFunc}, our methods treat any loss function as a black-box and retrieve derivatives via sampling.
Comparing the performance of \method{OurHVP} and  \method{OurHVPA} across a task with two and ten dimensions, we find the expected relation: \method{OurHVP} is faster in low dimensions\refFact{HVPFasterForLowDimension} than in higher dimensions for a similar task\refFact{HVPSlowerForHighDimension}.

We notice that the \method{OurHVPA} is not out-performing the \method{OurHVP} and \method{OurH} in the \task{Mug} task\refFact{HVPA1DIM}. This is because there is a single scene parameter for this task, thus, the higher order derivative is all in order of one.
With a higher variance for \method{OurHVPA}, it could be a little slower.
The effect of using the aggregate starts to show from \task{Shad}, where \method{OurHVPA} reaches the $1:1000$ convergence the fastest\refFact{HVPABestForShad}.

For all the rendering tasks, only the \task{Bunny} task converges for \method{Mitsuba}\refFact{MitsubaConvergesOnBunny}. Since the other tasks have plateaus in the loss function, the analytical gradient from inverse rendering is effectively zero, so it would be hard to converge within a reasonable time. The log-log plot shows in addition that derivatives from the inverse rendering were slower than the sampled ones. On the contrary, our methods not only smooth out the plateaus, but also improve the speed of the derivative computation. 

The drawback of constructing full Hessians is seen in the \task{Texture} task, where \method{OurH} is more than 40 times slower than the \ac{HVP} due to the matrix size\refFact{HBadInHighDimensions}.
For the most demanding task, \task{CNN} and \task{CNN5}, and even our best method\refFact{OurHVPANotOnCNN} is not able to reduce the error below $1:1000$.
At the $1:1000$-level, the more elaborate differentials\refFact{OurHVPNotFasterOnCNN} is not faster than the basic one, but faster than first-order\refFact{FR22SlowerOnCNN}.
Finally, the advantage of our ability to compute \acp{HVP} is seen when comparing \task{CNN} and \task{CNN5}: while a CNN producing a single parameter is doable with Hessians\refFact{OurHVPNotFasterOnCNN}, but if we go to higher dimensions, the aggregate strategy pays off again and is markedly faster\refFact{OurHVPSlowerForCNN5}.

In conclusion, albeit convergence varies with problem characteristics and dimensionality, we find that our best method, \method{OurHVPA} converges most reliably, beating the other algorithms by an average factor of 2.71$\times$ in wall-clock units. 
We show examples of task initialization and outcome, and additional evaluation in the supplemental. 

\mysection{Limitations and Conclusion}{conclusion}

\noindent \textbf{Limitations.} From the above experiments, it becomes apparent that our method performs well on lower-dimensional examples, but that its performance delta lessens with increasing problem dimensionality (see e.g. \task{Texture}). 
We hypothesize that this is due to sparser sampling in higher dimensions and exacerbated MC noise and acknowledge that further research in this direction is needed. 
We include an analysis on how our method degrades under MC rendering noise in the supplemental.

\noindent \textbf{Conclusion.} The availability of second-order gradient information gives rise to an exciting avenue of future research in inverse rendering. 
While these algorithms are not new, their use in vision \& graphics remains limited due to the convenience and widespread adoption of first-order gradient descent and the increased per-iteration cost that is usually associated with higher-order methods.
In this work, our efficient estimators have shown the latter to be negligible in real-world optimization scenarios.
We hope that this will inspire future research into unbiased, low-variance estimators of higher-order optimization methods.
%(\eg, the third-order Halley's method, which would enable cubic convergence). 

% \pagebreak

\setlength{\bibsep}{0.0pt}
{\small
\bibliographystyle{plainnat}
\bibliography{main}
}

\pagebreak

\appendix

\clearpage
\mysection{Gradient derivation}{gradient_derivation}
The $i$-th element of the gradient of the Gaussian, is
\begin{align}
\nabla_i\normal(\offsets)
&=
\frac{\partial_i\normal(\offsets)}{\partial \offset_i}
\\
&=
\frac{\mathrm d\normal(\offset_i)}{\mathrm d\offset_i}
\prod_{\substack{j=1\\j\neq i}}^\dimension
\normal(\offset_j)\\
&=
-\frac{\offset_i}{\bandwidth^2}
\normal(\offset_i)
\prod_{\substack{j=1\\j\neq i}}^\dimension
\normal(\offset_j)\\
&=
-\frac{\offset_i}{\bandwidth^2}
\normal(\offsets),
\end{align}
where we use the overloaded convention that $\normal(\offsets)$ takes a vector \offsets and $\normal(\offset_i)$ the $i$-th element, a scalar $\offset_i$, to produce the one or \dimension-dimensional Gaussian. 

This differs from \citet{fischer2023plateau} who only blur in the direction in which they differentiate, as in 
\begin{equation}
\nabla_i\normal(\offset_i)
=
\frac{\offset_i}{\bandwidth^2}
\normal(\offset_i),
\end{equation}
while it is more consistent with higher-order differentials to blur all dimensions.

\mysection{Smooth Gradient Marginalization}{gradient_sampling_derivation}
The sampling of 1D Gaussian gradient was derived by \citet{fischer2023plateau} who constructed a PDF \gradientPDF.
For a $n$-dimensional Gaussian gradient, we need to marginalize and sample the dimensions individually.
Marginalization would be integration over all dimensions $j\neq i$, except the one we look for $i$, so:
\begin{align}
&
\hphantom{\frac1Z}
\int_{\offsets_{j \neq i}}
\pdf^\mathrm G(\offsets)
\mathrm d\offsets_{j \neq i}
.\\
\intertext{Writing out the CDF, a positivized and scaled version of the PDF \pdf, where $1/Z$ is the partition function, gives}
&
\frac1Z
\int_{\offsets_{j \neq i}}
\frac12
|\nabla_i\normal(\offsets)|
\mathrm d\offsets_{j \neq i}=
\\
&
\frac1{Z}
\int_{\offsets_{j \neq i}}
|\nabla_i\normal(\offsets)|
\mathrm d\offsets_{j \neq i}
.\\
\intertext{Writing the $n$-D Gaussian as product of $n$ 1D Gaussians}&
\frac1{Z}
\int_{\offsets_{j \neq i}}
|\nabla_i
\prod^N_{j=1}
\normal(\offset_j)|
\mathrm d\offset_{j \neq i}
.\\
\intertext{As we differentiate only by $\offset_i$, all other factors are 1, so}
&\frac1{Z}
\int_{\offset_{j \neq i}}
|\nabla_i\normal(\offset_i)|
\mathrm d\offset_{j \neq i}
.\\
\intertext{As we are integrating over all $\offset_{j\neq 1}$, integration becomes multiplication with the domain's measure, 1.}&
|\nabla_i\normal(\offset_i)|
.
\end{align}

\mysection{Hessian derivation}{hessian_derivation}

The diagonal elements of the Hessian of the Gaussian are
\begin{align}
\nabla^2\normal(\offsets)_{ii}
&=
\frac{\partial^2_i\normal(\offsets)}{\partial^2\offset_i}
\\
&=
\frac
{\partial}
{\partial\offset_i}
\left(
-
\frac{\offset_i}{\bandwidth^2}
\normal(\offset_i)
\prod_{\substack{j=1\\j\neq i}}^\dimension
\normal(\offset_j)
\right)
\\
&=
\frac
{\partial}
{\partial\offset_i}
\left(
-
\frac{\offset_i}{\bandwidth^2}
\normal(\offset_i)
\right)
\prod_{\substack{j=1\\j\neq i}}^\dimension
\normal(\offset_j)
\\
&=
\left(
-\frac{1}{\bandwidth^2}+
\frac{\offset_i^2}{\bandwidth^4}
\right)
\prod_{\substack{j=1\\j\neq i}}^\dimension
\normal(\offset_j)
\\
&=
\left(
-\frac{1}{\bandwidth^2}+
\frac{\offset_i^2}{\bandwidth^4}
\right)
\normal(\offsets)
.
\end{align}

The non-diagonals of the Hessian of the Gaussian are
\begin{align}
\nabla^2\normal(\offsets)_{ij}
&=
\frac
{\partial^2 \normal(\offsets)}{\partial_i\offset_i\partial_j\offset_j}
\\
&=
\frac
{\partial}
{\partial\offset_j}
\left(
-
\frac{\offset_i}{\bandwidth^2}
\normal(\offset_i)
\prod_{\substack{k=1\\k\neq i}}^\dimension
\normal(\offsets_k)
\right)
\\
&=
\frac{\offset_i}{\bandwidth^2}
\normal(\offset_i)
\frac{\offset_j}{\bandwidth^2}
\normal(\offset_j)
\prod_{\substack{k=1\\k\neq i,j}}^\dimension
\normal(\offsets_k)
\\
&=
\frac{\offset_i}{\bandwidth^2}
\frac{\offset_j}{\bandwidth^2}
\normal(\offset_i)
\normal(\offset_j)
\prod_{\substack{k=1\\k\neq i,j}}^\dimension
\normal(\offsets_k)
\\
&=
\frac{\offset_i\offset_j}{\bandwidth^4}
\normal(\offsets)
.
\end{align}

A remark: It might appear, that diagonal is a special case of off-diagonal, but for differentiation, that is not true, as on the diagonal, the variable we differentiate in respect to appears twice, in the sense that $\mathrm d uv/\mathrm d u=v$ and $\mathrm d uv/\mathrm d v=u$, but $\mathrm d uu/\mathrm d u=2u$.

\mymath{\outerFunction}f
\mymath{\innerFunction}g
\mymath{\innerArgument}{\mathbf x}
\mymath{\outerArgument}{\mathbf y}
\mymath{\outerValue}{\mathbf z}

\mycfigure{DetailDerivation}{Detailed plots of the functions involved in the derivation of the CDF for the diagonal elements of  $\nabla^2\normal$, extending \refFig{Derivation}.}

\mysection{Sampling diagonal of Hessian}{sampling_hessian}

Similar to \refSec{gradient_sampling_derivation}, for the diagonal of the Hessian, we can sample each dimension independently.
Thus, we can first derive the valid distribution of the second-order derivative of the one-dimensional Gaussian by positivization and scaling, and it will apply to higher dimensions:
The one-dimensional Gaussian's second-order derivative is 
\begin{align}
\label{eq:secondDerivativeGaussian}
\left(
-\frac{1}{\bandwidth^2}+
\frac{\offset_i^2}{\bandwidth^4}
\right)
\normal(\offset_i)
\end{align}

The roots of \refEq{secondDerivativeGaussian} are the $\offset_i$ for which
\begin{align}
\left(
-\frac{1}{\bandwidth^2}+
\frac{\offset_i^2}{\bandwidth^4}
\right)
\normal(\offset_i)
&=0
.
\intertext{As $\normal(\offset_i)>0$ for all $\offset_i$, the product can be 0 only if}
-\frac{1}{\bandwidth^2}+
\frac{\offset_i^2}{\bandwidth^4}
&=0 \qquad\text{and hence}
\\
\offset_i
&=
\pm\bandwidth
.
\end{align}
The function value between $-\bandwidth < \offset_i \leq \bandwidth$ is negative and hence needs to be positivised.
Since the second-order derivative should integrate to the gradient of the Gaussian, we know that it reaches zero as $\offset_i$ reaches infinity.
In conjunction with the fact that the second-order derivative is symmetric about the $y$-axis, we can conclude that the integral of the interval $-\bandwidth < \offset_i \leq \bandwidth$ should be twice the size of the integral of $\offset_i \leq -\bandwidth = \offset_i > \bandwidth$.
Thus, after positivization, the CDF should be scaled, such that it is $\frac14$ at $\offset_i = -\bandwidth$.
Solving for these equalities, we can get:
\begin{align}
    \beta \nabla\normal(-\bandwidth) 
    &=
    \frac14
    \\
    \beta 
    &=
    \frac1{4\nabla\normal(-\bandwidth)}
    .
\end{align}

So, for the positivised rescaled second-order derivative as a PDF of the distribution:
\begin{align}
\pdf^\mathrm H_{ii}
&=
|\beta \nabla^2\normal(\offset_i)|
.
\end{align}

We can get the integrating constant by flipping and translating the scaled gradient of Gaussian to arrive at the CDF function for the intervals:

\newcommand{\coolStart}[2]
{\quad\mathclap{#1#2}\hphantom{\frac12}}

\begin{equation}
\cdf^\mathrm H_{ii}(\offset_i)=
\begin{dcases}
    \coolStart{}{}
    \beta\nabla\normal(\offset_i)
    &
    \text{if }\offset_i<-\bandwidth,\\
    \coolStart{\frac12}+
    \beta\nabla\normal(\offset_i)&
    \text{if } 
    \offset_i \in \left[-\bandwidth, \bandwidth \right]
    \\
    \coolStart1-
    \beta\nabla\normal(\offset_i)&
    \text{if }\offset_i>\bandwidth.\\
\end{dcases}
\end{equation}

\mysection{Grey-box differentials}{combination}
Sometimes, differentials are in respect to a function that is a composition $\outerValue=\outerFunction(\outerArgument=\innerFunction(\innerArgument))$ of an inner function with known analytic differentials \innerFunction (white box) and an outer function \outerFunction with differentials that need to be sampled (black box).
For first order (gradient), this is
\[
\nabla_\outerValue
(
\outerValue=
\outerFunction(\innerFunction(\innerArgument))
)
=
\left(
\nabla_\innerArgument
\innerFunction(\innerArgument)
\right)^\mathsf T
\cdot
\nabla_\outerArgument
\outerFunction(
\outerArgument=
\innerFunction(\innerArgument))
,
\]
which means to take the Jacobian (as both \innerFunction and \outerFunction in general are vector-valued) of the inner function \innerFunction in respect to the inner argument \innerArgument and vector-matrix multiply this with the gradient of the outer function \outerFunction but in respect to the outer argument \outerArgument.
For the second order it is
\[
\nabla^2_\outerValue
\outerFunction(\innerFunction(\innerArgument))
\approx
\left(
\nabla_\innerArgument
\innerFunction(\innerArgument)
\right)^\mathsf T
\cdot
\nabla^2_\outerArgument
\outerFunction(\innerFunction(\innerArgument))
\cdot
\nabla_\innerArgument
\innerFunction(\innerArgument)
\]
which means again to take the gradient of the inner function, but multiply it with the Hessian, instead of the Jacobian of the composition in respect to the outer argument \cite{martens2020new}.

The aim of this exercise is to have the sampled gradients handle only the black-box part and the analytic gradients handle the non-sampled parts.
As the analytic parts are typically large (\eg in the order of the size of a neural network) compared to the number of physical rendering parameters (placement of light, cameras or objects), this can provide substantial advantages.

\mysection{BFGS/LBFGS method}{LBFGS}
Quasi-Newton methods are also a way to utilize the second order information for optimization, however, they approximate this information with zero or first order information. We tested the family of algorithms from the quasi-Newton methods that is known to be most effective, the BFGS algorithms \cite{sherman1949adjustment}. For this family of algorithms, the vanilla BFGS algorithm \cite{shanno1970conditioning}, along with BFGS with Armijo-Wolfe line search \cite{nocedal1999numerical}, and LBFGS \cite{nocedal1999numerical}, were tested, but they only converge for the \task{Quad} task. This is probably due to the noisy nature of the derivative estimation. To this end, damped BFGS \cite{dampedBFGS1} and adaptive finite difference BFGS \cite{berahas2019derivative} were also added, but neither changed the convergence of other tasks.

\mysection{Comparing to analytical Hessians}{Analytical}
To further validate our approach, we investigate a task where the analytic Hessians are available: the optimization of a sphere's Phong BRDF (7 unknowns) under point illumination.
We use this task to evaluate both analytical and sampled derivatives, as well as the corresponding first and second-order methods that employ them in \cref{fig:Material}. 
The dashed lines indicate the first‑order methods, while the solid lines represent the second‑order methods. 
% \begin{wrapfigure}[13]{r}{0.5\linewidth}
% \vspace{-7mm}
% \begin{center}
% \includegraphics[width=\linewidth]{images/Material.ai}
% \vspace{-5mm}
% \caption{BRDF optimization.}
% \label{fig:Material}
% \end{center}
% \end{wrapfigure}
\myfigure{Material}{BRDF optimization.}
We see that the sampled method is less accurate and takes longer due to the sample size and the bandwidth of sampling. 
The second‑order methods outperform the first‑order ones in both respects.
We show the outcome of our second-order optimization denoted as 'Ours' in \cref{fig:Material}. 

\mysection{Positive-Definite Hessians}{PSD}
Naive Newton will fail for cases where the problem's Hessian is not positive semi-definite (PSD). For a non-positive definite Hessian, there may exist a unbounded, negative eigenvalue. This means that when the Hessian is applied to a gradient vector, it may flip the gradient vector and cause opposite divergence from the gradient's descent direction. 
Although optimization is separate from our contribution of stochastic gradient estimation, we demonstrate how our method deals with this issue with an example. 
To this end, consider a negated 2D Gaussian $-\mathcal{N}(0, \sigma_1)$ whose Hessian is non-PSD everywhere. 
Its convolution with a second Gaussian $-\mathcal{N}(0, \sigma_2)$ results in a third Gaussian $-\mathcal{N}(0, \sigma_3)$, whose Hessian is also not PSD everywhere. 
In this example, we know the analytic expressions of all the Gaussians, so we can compute their Hessians and compare them to our estimates.
% \begin{wrapfigure}[13]{r}{0.5\linewidth}
%     \centering
%     \includegraphics[width=\linewidth]{images/NegatedGaussian.ai}
%     \caption{Optimization on a problem with non-PSD Hessians.}
%     \label{fig:NegatedGaussian}
% \end{wrapfigure}
\myfigure{NegatedGaussian}{Optimization on a problem with non-PSD Hessians.}
The mean error of our method across the interval $(-3,3)^2$ is within $1\times10^{-4}$ for 100 samples.
With Hessian modification and trust region, running an optimization finds the correct minimum at (0,0) for any starting point in $(-3,3)^2$. 
\cref{fig:NegatedGaussian} shows the optimization error across an ensemble of 20 runs in addition to a top-view of the optimization trajectories for both the analytic and sampled derivatives. 
This indicates that, for the right combination of estimator and optimizer, non-PSD is not necessarily a problem. 

\mycfigure{overview_init_ours}{Overview of all tasks we study (columns): the first row shows the task initialization, the starting point of the optimization. The middle row shows the outcome of optimizing with our estimated Hessians with \method{OursHVPA}, while the bottom row shows the ground truth for each task.}

\mysection{Robustness analysis}{Robustness}
We investigate our method's robustness to:

\noindent \textbf{Initialization}.
All our experiments are performed over an ensemble of 20 random starting points. 
The reported numbers and findings are the average of these optimization outcomes. 

\noindent \textbf{Hessian sample count.}
We repeated the \task{Mug} task with different numbers of samples used to estimate the Hessian during optimization (using multiples of two, due to antithetic sampling).
Our overall findings are consistent with prior work and indicate that increasing the number of estimation samples does help convergence. 
However, the increase timed for higher sample counts is not offset by improved convergence for this task, as the per-iteration convergence gain does not sufficiently compensate for the slowdown (left subplot in \refFig{robustness}).
This indicates that the minimal number of antithetic samples (two) can be optimal for a relevant task.

\noindent \textbf{Rendering MC noise.}
We have repeated the \task{Mug} experiment with 0.25$\times$, 0.5$\times$, 1$\times$, 2$\times$, and 3$\times$ the number of rendering samples and did not observe significant change in optimization quality and convergence.
However, as the sample count for rendering increased, the time taken to render each image is longer, leading to a slower convergence rate for higher sample counts (center subplot in \refFig{robustness}).
This provides us with a data point that noise from low-spp MC rendering is not a dominating limiting factor for our method, although a more in-depth investigation would be needed to reliably confirm this across all experiments. 

\noindent \textbf{Real-world compression noise.} 
We repeated the \task{Mug} task with JPEG compression at the 2\%, 4\%, 6\%, 9\%, and 14\% levels and did not observe significant degradation in convergence (right subplot in \refFig{robustness}).
The difference in the final result is caused by the loss calculation between the rendered image and the noisy, compressed image.

\myfigure{robustness}{Robustness evaluation (see text). All plots are displayed on a log-error vs. log-time scale.}

\mysection{How much variance is reduced?}{VarianceAnalysis}
In \refFig{Variance}, we perform a variance analysis of the different estimators on different differentiable quantities.
We see that all estimators converge linearly in a log-log plot.
The optimal estimator (dotted) has the lowest variance and hence would lead to the least noise in optimization, but at the expense of evaluating quadratically many elements for Hessians.
Our aggregate sampling (solid line) performs slightly worse, but much, better than uniform sampling (thin line) would do.
\myfigure{Variance}{Estimator variance (lines) for different operators (plots).}

\begin{table*}[h]
\small
\begin{tabular}{@{}lllllllllll@{}}
\toprule
Method   & Parameter     & \task{Quad} & \task{Box2} & \task{Box10} & \task{Mug}  & \task{Shad} & \task{Bunny} & \task{Texture} & \task{CNN}  & \task{CNN5} \\ \midrule
          & Samples & 4    & 6    & 6     & 1    & 1    & 2     & 4       & 1    & 1    \\
          & Sigma (start)    & 1    & 1.5  & 0.6   & 3    & 0.5  & 1     & 0.5     & 0.5  & 0.5  \\ 
          \midrule
\method{FR22}      & Learning rate   & 0.5  & 0.3  & 0.05  & 0.1  & 0.02 & 0.02  & 0.05    & 1e-4 & 1e-4 \\
          & Sigma (end)      & 0.01 & 0.01 & 0.1   & 0.01 & 0.01 & 0.01  & 0.1     & 0.01 & 0.01 \\
          \midrule
\method{OurG}      & Learning rate   & 0.5  & 0.3  & 0.05  & 0.1  & 0.02 & 0.02  & 0.05    & 1e-4 & 1e-4 \\
          & Sigma (end)      & 0.01 & 0.01 & 0.1   & 0.01 & 0.01 & 0.01  & 0.1     & 0.01 & 0.01 \\
          \midrule
\method{OurH}      & Trust region        & 50   & 2    & 10    & 3    & 3    & 2     & 10      & 1e-2 & 1e-2 \\
          & Sigma (end)      & 0.01 & 0.01 & 0.1   & 0.01 & 0.01 & 0.005 & 0.1     & 0.01 & 0.01 \\
          & Line search iteration    & 5    & 10   & 3     & 1    & 1    & 1     & 5       & 2    & 2    \\
          & Line search tolerance   & 1e-3 & 1e-3 & 1e-3  & 1e-3 & 1e-3 & 1e-3  & 1e-3    & 1e-3 & 1e-3 \\
          & Recompute           & 5    & 10   & 30    & 10   & 10   & 5     & 30      & 20   & 20   \\
          \midrule
\method{OurHVP}    & Trust region        & 50   & 2    & 10    & 4    & 5    & 4     & 10      & 1e-2 & 1e-2 \\
          & Sigma (end)      & 0.05 & 0.01 & 0.1   & 0.01 & 0.01 & 0.01  & 0.1     & 0.01 & 0.01 \\
          & Line search iteration    & 5    & 3    & 3     & 2    & 1    & 1     & 5       & 2    & 2    \\
          & Line search tolerance   & 1e-3 & 1e-3 & 1e-3  & 1e-3 & 1e-3 & 1e-3  & 1e-3    & 1e-3 & 1e-3 \\
          & Recompute           & 5    & 10   & 30    & 2    & 5    & 10    & 30      & 20   & 20   \\
          \midrule
\method{OurHVPA} & Trust region        & 50   & 2    & 10    & 4    & 3    & 4     & 10      & 1e-2 & 1e-2 \\
          & Sigma (end)      & 0.05 & 0.01 & 0.1   & 0.01 & 0.01 & 0.01  & 0.1     & 0.01 & 0.01 \\
          & Line search iteration    & 5    & 1 0  & 3     & 1    & 10   & 2     & 5       & 2    & 2    \\
          & Line search tolerance   & 1e-3 & 1e-3 & 1e-3  & 1e-3 & 1e-3 & 1e-3  & 1e-3    & 1e-3 & 1e-3 \\
          & Recompute           & 5    & 10   & 30    & 5    & 5    & 2     & 30      & 20   & 20  \\
          \bottomrule
\end{tabular}
\caption{Hyperparameters for our methods (rows) on the different tasks (columns). All parameters, including those of our competitors, have been optimally chosen.}
\label{tab:hyperparameters1}
\end{table*}
\mysection{Additional Experiments}{AdditionalExperiments}
We perform two additional experiments to show that our method can successfully differentiate diverse light transport scenarios: in \refFig{caustic} we have repeated the caustic example from ZeroGrads \cite{fischer2024zerograds}, where the goal is to optimize a heightfield, parametrized by a 1,024-dimensional B-spline, such that the caustic it creates matches a reference image. 
Our method performs well on this task, even thought the loss landscape is highly non-convex and the optimization variables exhibit highly non-local image-space interactions. 
\myfigure{caustic}{Inverse optimization of a heightfield such that the caustic it creates when light shines through it matches a reference. CMA-ES does not converge on this task.}
\myfigure{mirror_bunny}{Two-bounce mirror optimization experiment.}
Additionally, we optimize a two-bounce experiment, where the rotation (around the up-axis) and translation (along x,y) of the Stanford bunny are optimized, while the bunny is only observed through two mirrors. 
The left subfigure in \refFig{mirror_bunny} shows the experiment. 
We use our method \method{OursHVPA} and observe a successful optimization outcome.

\mysection{Hyperparameters}{Hyperparameters}
In this section, we detail the hyperparameters of our experiments and show the initial-, output- and ground-truth images for each task. 
The hyperparameters are shown in \cref{tab:hyperparameters1}, and the task images are shown in \refFig{overview_init_ours}.
The sample size, which is always antithetic and doubles the sample size, applies to our methods and FR22. 
All Sigma annealing, which controls the bandwidth \bandwidth of the distribution \cite{fischer2023plateau}, is scheduled linearly and has a start and end value. 
For first-order methods, the tunable hyperparameter is the learning rate. 
For second-order ones, the trust region value shows the initial search bound, and recompute is the number of iterations until the line search is re-estimated.

\end{document}

%% file: preamble.tex
\usepackage{soul}
\usepackage{xcolor}

%% file: tr-commands.tex
\usepackage{soul}
\usepackage{amsmath}
\usepackage{amssymb}
\usepackage{wrapfig}
\usepackage{booktabs}
\usepackage{multirow}
\usepackage{xspace}
\usepackage{xcolor}

\def\figurePath{images/}

\NewDocumentCommand{\rot}{O{45}
O{1em}m}{\makebox[#2][l]{\rotatebox{#1}{#3}}}%

\def\myfigure#1#2{%
    \begin{figure}[htb]%
    \centering\includegraphics*[width = \linewidth]{\figurePath#1}%
    \vspace{-.2cm}%
    \caption{#2}%
    \vspace{-.2cm}%
    \label{fig:#1}%
    \end{figure}%
}

\def\mycfigure#1#2{%
    \begin{figure*}[htb]%
    \centering\includegraphics*[width = \linewidth]{\figurePath#1}%
    \vspace{-.2cm}%
    \caption{#2}%
    \vspace{-.2cm}%
    \label{fig:#1}%
    \end{figure*}%
}

\newcommand{\refSec}[1]{Sec.~\ref{sec:#1}}
\newcommand{\refFig}[1]{Fig.~\ref{fig:#1}}
\newcommand{\refEq}[1]{Eq.~\ref{eq:#1}}
\newcommand{\refTab}[1]{Tab.~\ref{tab:#1}}
\newcommand{\refAlg}[1]{Alg.~\ref{alg:#1}}

\soulregister\ref7
\soulregister\cite7
\soulregister\refFig7
\soulregister\refAlg7
\soulregister\refSec7
\soulregister\cite7
\soulregister\ref7
\soulregister\pageref7
\soulregister\shortcite7
\soulregister\eg0
\soulregister\ie0
\soulregister\etal0
\soulregister\task7

\DeclareGraphicsExtensions{.png,.jpg,.pdf,.ai,.psd}
\newcommand{\mysection}[2]{\vspace{-.1cm}\section{#1}\label{sec:#2}\vspace{-.1cm}}
\newcommand{\mysubsection}[2]{\subsection{#1}\label{sec:#2}\vspace{-.1cm}}

\definecollection{mymaths}
\newcommand{\mymath}[2]{
    \newcommand{#1}{\TextOrMath{$#2$\xspace}{#2}}
    \begin{collect}{mymaths}{}{}{}{}
    #1
    \end{collect}
}

\definecolor{colorA}{HTML}{4285f4}
\definecolor{colorB}{HTML}{ea4335}
\definecolor{colorC}{HTML}{fbbc04}
\definecolor{colorD}{HTML}{34a853}
\definecolor{colorE}{HTML}{ff6d01}
\definecolor{colorF}{HTML}{46bdc6}
\definecolor{colorG}{HTML}{000000}
\definecolor{colorH}{HTML}{777777}
\definecolor{colorI}{HTML}{bdd6ff}
\definecolor{colorJ}{HTML}{6a9e6f}

\usepackage{pifont}

\newcounter{fact}
\newcommand{\fact}[1]{%
 \refstepcounter{fact}%
 \label{fact:#1}%
 \rlap{\textsuperscript{\arabic{fact}}}%
}
\newcommand{\refFact}[1]{\textsuperscript{\ref{fact:#1}}}